\documentclass[9pt,twocolumn,twoside]{osajnl}

\journal{ol} 

\setboolean{shortarticle}{true}

\title{Non-Hermitian spectral changes in the scattering of partially coherent radiation by periodic structures}

\author[1,*]{P. A. Brand\~ao}
\author[1]{S. B. Cavalcanti}

\affil[1]{Universidade Federal de Alagoas, Instituto de F\'isica, 57072-900, Brazil}

\affil[*]{Corresponding author: paulo.brandao@fis.ufal.br}

\begin{abstract}
The physical aspects of partially coherent radiation interacting with deterministic non-Hermitian periodic materials remain largely unexplored in the statistical optics literature. Here, we consider the scattering of partially coherent radiation by a deterministic periodic medium, symmetric under the simultaneous transformations of parity inversion and time reversal, that is, a parity-time (PT)-symmetric periodic medium. Taking into account light fluctuations, one is able to describe the spectrum changes on propagation and the influence of the coherence-driven angular divergence effect. The far-field spectral density profile is found to depend crucially on the loss/gain properties of the material, giving rise to unexpected and contrasting spectral diffraction profiles when compared to the Hermitian ones. 
\end{abstract}

\setboolean{displaycopyright}{true}

\begin{document}

\maketitle

A non-Hermitian formulation of quantum mechanics has been initiated by Bender and Boettcher in 1998 \cite{Bender1998a,Bender2002,Bender1999a,Bender2007a,bender2018pt}. Nowadays, is has been well established, that a Hamiltonian which is invariant under the combined action of parity inversion and time reversal symmetries (PT symmetry), may have all-real valued eigenvalues.  This is a necessary but not a sufficient condition 
to guarantee the reality of the spectrum \cite{Bender1998a}. 
In a series of interesting papers, Mostafazadeh has shown that operators having PT symmetry belong to the more general class of pseudo-Hermitian operators \cite{Mostafazadeh2002c,Mostafazadeh2002d,Mostafazadeh2002e}. Regarding experimental results on PT-symmetric systems, optical physics has provided the most accessible route to verify the reality of eigenvalues of PT-symmetric operators along with the symmetry breaking transition point present in non-Hermitian configurations. This was possible due to the  analogy between the Schr\"odinger and the paraxial wave equation \cite{Longhi2017a,Ruschhaupt2005a,Ruter2010}, where the potential function is interpreted as the refractive index. In the last 
decades, non-Hermitian PT-optical lattices have been extensively 
investigated exhibiting exotic phenomena such as: non-unitary evolution \cite{Berry_2008}, unidirectional invisibility 
\cite{PhysRevLett.106.213901}, which was later experimentally evidenced in PT-symmetric syntethic photonic lattices \cite{nature_12} and in a PT-symmetric metamaterial \cite{nature_13}. Spectral singularities of Bragg 
scattering processes 
in periodic optical lattices have been associated with 
exceptional points, exhibiting a secular growth of plane waves that satisfy the Bragg condition, which saturates in case of wave packets with a broad momentum distribution \cite{Longhi2010,PhysRevAaeva}. The effect of broadband quasi-PT symmetry in a 
finite frequency range on pulse propagation, has also been investigated in \cite{PhysRevA.98.053844}.  The interest in 
 non-Hermitian structures 
 has grown steadily and, in the last few years various aspects of scattering periodic structures were studied, such as the transmission properties of finite periodic structures \cite{PhysRevLett.119.243904} as well as
 resonant phenomena in layered periodic ones \cite{Shramkova_2016}. 
 
However, up to now all studies on PT-symmetric systems  do not take into account fluctuations, could that be of the electromagnetic field or the medium, or both. As pointed out by Wolf \cite{Wolf2013a}, who studied the scattering of a partially coherent radiation field by a periodic medium, short wavelength radiation is not complete spatially coherent and therefore its  random nature must be taken into account in diffraction studies. Some aspects of diffracted wavefields by non-Hermitian gratings were recently published in the literature \cite{Gao2018a,Shui2018,Zhu2016a}. However, to achieve some insight about the dynamical behavior of some observables such as the spectral density one must consider the coherence properties of the radiation field, particularly if one is dealing with short wavelength radiation. Within the limits of the Born approximation, the scattering of  partially coherent radiation from a random medium, that is, a medium whose dielectric function is a random function of both position and time, has been reported to generate frequency shifts \cite{PhysRevLett.63.2220}. Later on, a study on the diffraction of partially coherent beams from three dimensional periodic structures has been addressed, demonstrating coherence-induced angular shifts of the diffraction orders due to the finite size of the coherence area \cite{PhysRevE.52.6833}. Therefore, inspired by the elegant theory of coherence, developed mainly by Wolf, based on stationary random processes \cite{wolf2007introduction}, and considering that PT-symmetric structures have been artificially obtained and have been the object of many studies worldwide, in the following we have studied the scattering of a partially coherent radiation field by a periodic PT-symmetric medium, within the first-order Born approximation. In contrast with the Hermitian potential, we find radical changes in the spectra profiles due to the combined effect of PT symmetry and partial coherence on propagation.

Let us begin by representing light as a stationary random process, characterized by the spectral degree of coherence $\mu^{(i)}(\mathbf{r}_1,\mathbf{r}_2;\omega)$ \cite{wolf2007introduction}. Next, we suppose that it is diffracted by a periodic medium described by the function  $V(\mathbf{r}) = V(\mathbf{r+ a})$, which is  proportional to the refractive index of the material. After the interaction, in the asymptotic limit, the spectral density of the scattered radiation is given by \cite{wolf2007introduction}
\begin{equation}\label{sinf}
\begin{split}
	S^{(\infty)}(\mathbf{r};\omega) &= \frac{S^{(i)}(\omega)}{r^2} \\
	& \times\int\int _D d^3 r_1 d^3 r_2\mu^{(i)}(\mathbf{r}_1,\mathbf{r}_2;\omega)V^{*}(\mathbf{r}_1) \\
	&\times V(\mathbf{r}_2)\exp[-ik \textbf{s} \cdot (\textbf{r}_2 - \textbf{r}_1)],
\end{split}
\end{equation}
where $k = \omega/c$ with $c$ being the speed of light in vacuum and the first-order Born approximation was used. The unit vector $\textbf{s}$ is directed from the scatterer to the observation point. Let us assume a one-dimensional periodic medium described by $V(\textbf{r}) = v(x)\delta(y)\delta(z)$ where $v(x)$ is a periodic function with period $a$ and $\delta$ denotes the Dirac delta function. Following \cite{Wolf2013a}, we also consider that the incident spectral degree of coherence has a Gaussian form given by
\begin{equation}
	\mu^{(i)}(\rho_1,\rho_2;\omega) = \exp\left[ - \frac{(\rho_2 - \rho_1)^{2}}{2\sigma^2} \right],
\end{equation}
with $\sigma$ denoting the range of transversal coherence with $(\rho_1,\rho_2)$ representing points at the cross section of the beam. Let us now choose a one dimensional PT-symmetric material by requiring that $(\mathcal{P} \mathcal{T})V(\textbf{r})(\mathcal{P} \mathcal{T})^{-1} = V(\textbf{r})$ which, in turn, implies that $v(x) = v^* (-x)$ or, in words, the real (imaginary) part of $v(x)$ must be even (odd) under the transformation $x \rightarrow -x$. The operators $\mathcal{P}$ and $\mathcal{T}$ are the parity inversion ($\bf{r}\rightarrow -\bf{r}$) and time reversal operators ($i\rightarrow -i$), respectively. Thus, to satisfy these conditions, we consider the PT-symmetric medium described by
\begin{equation}\label{potential}
	v(x) = \frac{1}{2} + v_r\cos\left( \frac{2\pi x}{a} \right) + iv_i \sin\left( \frac{2\pi x}{a} \right),
\end{equation} 
where $v_r$ and $v_i$ are real and positive numbers which may depend on the frequency $\omega$. The function $v(x)$ is a truncated Fourier series of the form 
\begin{equation}
	v(x) = \sum_{n=-\infty}^{\infty} c_n \exp\left( \frac{2\pi i x n}{a} \right)
\end{equation}
with coefficients $c_{n}$ given by $c_{\pm 1} = (v_r \pm v_i)/2$ and $c_0 = 1/2$, and $c_n = 0$ for $n\geq 2$. The coefficients $c_n$ are real numbers according to  the condition $v(x)=v^* (-x)$ which implies that $c_n=c_n^*$, as can be easily verified. In terms of $c_n$, \eqref{sinf} can be recast into
\begin{equation}\label{sinf2}
\begin{split}
	S^{(\infty)}(\mathbf{r};\omega) &= \frac{LS^{(i)}(\omega)}{r^2} \sum_{n=-\infty}^{\infty}|c_n|^2 \int_L dx' \exp\left[ -\frac{(x')^2}{2\sigma^2} \right] \\
	&\times  \exp[-i(k\textbf{s}\cdot \textbf{x})x']\exp\left( \frac{2\pi in x'}{a} \right),
\end{split}
\end{equation}
where $L$ is a characteristic (large) length of the structure, $x' = x_2-x_1$ and $\textbf{x}$ is the unit vector in the $x$ direction. In the particular case of the material described in \eqref{potential} this integral is readily evaluated in closed form
	\begin{equation}\label{result}
	\begin{split}
	S^{(\infty)}(\mathbf{r};\omega) &= \frac{L\sigma\sqrt{2\pi} S^{(i)}(\omega)}{4r^2}\left\{ \exp\left[  \frac{-(k\sigma\cos\theta)^2}{2} \right] \right.\\
	&+\left. (v_r-v_i)^2\exp\left[\frac{-\sigma^2(k\cos\theta + 2\pi/a)^2}{2}\right] \right. \\  
	&\left. + (v_r+v_i)^2\exp\left[ \frac{-\sigma^2(k\cos\theta - 2\pi/a)^2}{2} \right] \right\}.
	\end{split}
	\end{equation}
where $\theta = \arccos(\bf{s}\cdot\bf{x})$. In what follows, we will assume that the spectral density of the incident wavefield has the following Gaussian profile: 
\begin{equation}
    \label{si}
    S^{(i)}(\omega) = S_0 \exp\left[ -\frac{(\omega - \omega_0)^{2}}{2\delta^2} \right],
\end{equation}
where $\omega_0$ is the central frequency, $\delta$ the bandwidth and $S_0$ the spectrum amplitude. 

\begin{figure}[htbp]
\centering
\includegraphics[width=0.7\linewidth]{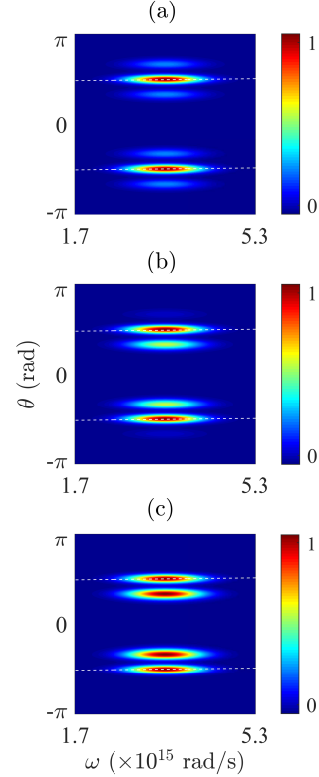}
\caption{(Color online) Normalized scattered spectral density of partially coherent radiation with $\sigma = 10\times(1/k)$. (a) Hermitian scattering with $v_i = 0$.  Non-Hermitian scattering with (b) $v_i = 0.25$ and (c) $v_i = 0.5$. The real part of the potential function is $v_r = 0.5$, the incident spectrum is described by \eqref{si} with $\omega_0 = 3.54\times10^{15}$ rad/s ($\lambda_0 = 532$ nm), $\delta = 0.1\omega_0$ and $a = 2\lambda$. The white dashed lines are fixed at $\theta = \pm \frac{\pi}{2}$.}
\label{fig1}
\end{figure}

To obtain sensible numerical estimates from this model, we assume that the incident spectral density is centered at $\omega_0 = 3.54\times 10^{15}$ rad/s ($\lambda_0 = 532$ nm) with the bandwidth $\delta = 0.1\omega_0$. Figure \ref{fig1} depicts the normalized spectral density $\tilde{S}(\theta,\omega) = 4r^2 S(\theta,\omega)/\sqrt{2\pi}\sigma L S_0$ in the $(\theta,\omega )$ plane for an incident light with a high degree of coherence, $\sigma = 10\times (1/k)$. Part (a) of Figure \ref{fig1} illustrates the Hermitian case where $v_i = 0$. In this situation, we recover the results studied in \cite{Wolf2013a} in which a symmetrical interference pattern is obtained with a central maximum at $\theta_c = \pm\pi/2$, highlighted by the two white dashed lines, accompanied by two secondary maxima at angles $\theta_{\text{ext}} = \pm 2\pi/3$ and $\theta_{\text{in}} = \pm \pi/3$. The subindex ext (in) indicate the outermost (innermost) maxima relative to $\theta = 0$. The value $v_r = 0.5$ is assumed in all plots. Notice that $\theta_c$ corresponds to a direction perpendicular to the lattice. Therefore, in this situation, we obtain the usual diffraction pattern present in crystallographic systems. If we now allow the lattice to absorb energy from and to give energy to the wavefield, by adjusting the non-Hermitian parameter to $v_i = 0.25$, the result is shown in part (b) of Figure \ref{fig1}. In this case, the amplitudes of the secondary maxima change according to: 
\begin{align}\label{amp}
\begin{split}
 &\text{Amplitudes at } \theta_{\text{in}} \sim (v_r + v_i)^2 ,
\\
 &\text{Amplitudes at } \theta_{\text{ext}} \sim (v_r - v_i)^2.
\end{split}
\end{align}
The central maximum at $\theta_c$ remains unchanged as we activate the loss/gain properties of the lattice. The effect of the non-Hermiticity in this model changes only the amplitude factors as shown in \eqref{amp}. Therefore, it is possible to adjust the spectral properties of the wavefield by tuning the loss/gain properties of the lattice. The situation shown in part (c) of Figure \ref{fig1} was obtained by letting $v_i = v_r$. In this case, the secondary maxima at $\theta_{\text{ext}}$ are zero. This scenario generates an asymmetrical spectral density in relation to the perpendicular direction of the lattice. In non-relativistic quantum mechanics described by a Schr\"odinger equation with potential \eqref{potential}, the situation $v_r = v_i$ represents the symmetry breaking point where the eigenvalues become degenerate. In the present model, the physical effect of $v_r = v_i$ is the disappearance of the spectral amplitude at $\theta_{\text{ext}}$ whose energy is transferred to the secondary maxima located at $\theta_{\text{in}}$, while the central maximum at $\theta_c$ remains unchanged. Furthermore, the enhanced maximum become as intense as the central one, so that the non-Hermitian spectrum becomes a twin pair of maxima, in both directions alike.  

\begin{figure}[ht]
\centering
\includegraphics[width=0.7\linewidth]{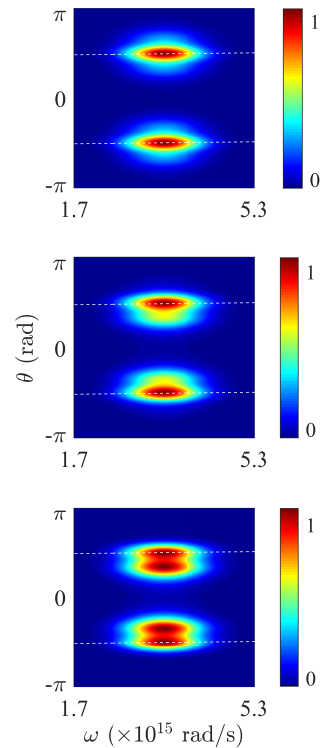}
\caption{(Color online) Same as Figure \ref{fig1} except for the wavefield, 
which now is assumed within the low-coherence regime with $\sigma = 5\times(1/k)$. (a) Hermitian and (b,c) complex lattice.}
\label{fig2}
\end{figure}

Let us now decrease the coherence length of the radiation field to $\sigma = 5\times (1/k)$. In this situation, the temporal and spatial variations of the wavefield observed in any two pair of points have smaller statistical correlations than the previous case. Figure \ref{fig2} shows the normalized spectral density for the same values used in Figure \ref{fig1}, except for the coherence length which is now $\sigma = 5\times (1/k)$. For the Hermitian case, shown in part (a), the two secondary maxima disappear and we no longer obtain the Hermitian interference pattern, only two maxima of equal intensities, at $\theta = \theta_c$. The situation is quite different when the lattice is equipped with gain and loss regions. Parts (b) and (c) of the same figure show the spectral density for the non-Hermitian case. It can be seen from these plots that two maxima are still discernible, even for an incident radiation field in the low coherence regime. Part (c), in particular, represents a  non-Hermitian interference pattern induced by the complex lattice. Lattice properties are generally obtained by examining the interference pattern they generate and, since in this case this pattern depends on the Hermiticity of the lattice, this result could be useful for effective spectroscopy and grating diffraction research.

\begin{figure}[ht]
\centering
\includegraphics[width=0.7\linewidth]{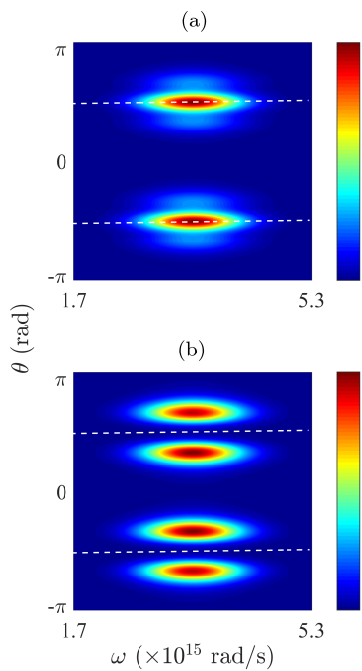}
\caption{(Color online) Normalized spectral density of the scattered wavefield in the regime where $v_i \gg v_r$. (a) Hermitian lattice with $v_i = 0$ and (b) non-Hermitian lattice with $v_i \gg v_r$ described by \eqref{viggvr}. The coherence length $\sigma$ is the same as in Figure \ref{fig2}.}
\label{fig3}
\end{figure}

As pointed out before, in the particular situation where $v_i = v_r$, the eigenvalues of the Schr\"odinger equation with the potential described by \eqref{potential} undergoes a symmetry breaking transition. In the context of 
optics and the paraxial wave equation, the propagation of a wavefield
also exhibits the process of energy transfer between normal modes. However, 
above this point, that is, when $v_i > v_r$, some of the eigenvalues may become complex and are characterized by an uncontrolled increase of the amplitude of the electric field during propagation \cite{Brandao2017}. In contrast, in the present model we see no indication of a symmetry breaking point characterized by a divergent behavior of the spectral density. It should be noted, however, that the present approach has little resemblance with the non-relativistic Schr\"odinger equation such that we do not expect all characteristics of the former system to apply here. In particular, we are observing the wavefield far away from the material, working with an asymptotic solution. Therefore, our theory might be extended to include cases where $v_i \gg v_r =0.5$ and, indeed, in this limiting situation the spectral density is given by
\begin{multline}\label{viggvr}
    S^{\infty}(\theta,\omega) \sim S^{(i)}(\omega)\frac{L\sigma\sqrt{2\pi}v_i^2 }{4r^2} \left\{ \exp\left[\frac{-\sigma^2(k\cos\theta + 2\pi/a)^2}{2}\right] \right. \\
    \left. + \exp\left[\frac{-\sigma^2(k\cos\theta - 2\pi/a)^2}{2}\right] \right\}.
\end{multline}
The interesting thing about this equation is that the central maximum at 
$\theta_c$ is now negligible compared to the secondary ones which are, in fact, the only relevant quantities. This produces a symmetrical interference pattern for moderate values of $\sigma$. To see the difference between Hermitian and non-Hermitian lattices in this approximation, Figure \ref{fig3} shows the normalized spectral density for $v_i = 0$ in part (a) and for $v_i \gg v_r$ in part (b) within the low-coherence regime where $\sigma = 5\times (1/k)$. A clear non-Hermitian diffraction pattern with two secondary maxima is still visible even in the case of a partially coherent incident field, when compared to the Hermitian case. In particular, a comparison between the results presented in part (c) of Figure \ref{fig2} and part (b) of Figure \ref{fig3} indicates that a visible interference pattern can be obtained by controlling the loss/gain properties of the lattice with $\sigma$ fixed.  All these results suggest the possibility to tune and to improve the visibility of the spectral density,  by adjusting gain/loss regions of the complex lattice without changing the geometry of the system. We should also mention that there is a small shift in the central frequency of the scattered spectral density compared to the initial frequency $\omega_0$ that is solely due to the intrinsic nature of the correlations existent in the wavefield \cite{PhysRevLett.63.2220}. This shift is, however, negligible within the range of parameters considered in the plots and therefore is not visible in these pictures. Since the current experimental research interest on non-Hermitian systems is to consider materials with modulated loss, it is important to verify what classical coherence theory says in this particular context. If one assumes that, for example, $v(x) = v_r+ iv_i\sin^2(\pi x/a)$, then the material has no gain, only loss \cite{berry1998diffraction}. A simple calculation shows that the complex coefficients $c_n$ in the Fourier expansion are $c_0 = v_r + iv_i/2$ and $c_{\pm1} = -iv_i/4$. It is easy to see that none of the above effects are realized in this pure lossy configuration, implying that gain may be a decisive factor for these effects to appear.

Summing up, considering the importance of the statistical fluctuations on the evolution of the spectra of short wavelength radiation, we have investigated the scattering of partially coherent electromagnetic fields by a periodic PT-symmetric medium. Assuming a stationary random process to represent the incident partially coherent field, and within the first-order Born approximation, we obtain the spectral density of the scattered radiation. Our results have provided us with new physical insights into a class of scattering process of a stationary partially coherent field from a deterministic PT-symmetric periodic medium. Experimentally, artificial optical structures with controlled gain and loss regions have been fully pursued, and one may study various types of potential functions as well, as various types of statistical processes.  In this way our results should be quite useful in the understanding of the diffraction patterns obtained by the scattering of beams in any state of partial coherence from any type of PT-symmetric media, not necessarily a periodic medium.  Work in this direction is in current progress and will be published elsewhere. 
	
The authors are indebted to the Brazilian agencies CNPq and FAPEAL for partial financial support.

\bibliography{sample}

\begin{thebibliography}{10}
\newcommand{\enquote}[1]{``#1''}

\bibitem{Bender1998a}
C.~M. Bender and S.~Boettcher, \enquote{{Real spectra in non-hermitian
  hamiltonians having PT symmetry},} {\protect\JournalTitle{Physical Review
  Letters}} \textbf{80}, 5243--5246 (1998).

\bibitem{Bender2002}
C.~M. Bender, D.~C. Brody, and H.~F. Jones, \enquote{{Complex Extension of
  Quantum Mechanics},} {\protect\JournalTitle{Physical Review Letters}}
  \textbf{89}, 1--4 (2002).

\bibitem{Bender1999a}
C.~M. Bender, S.~Boettcher, and P.~N. Meisinger, \enquote{{PT-symmetric quantum
  mechanics},} {\protect\JournalTitle{Journal of Mathematical Physics}}
  \textbf{40}, 2201--2229 (1999).

\bibitem{Bender2007a}
C.~M. Bender, \enquote{{Making sense of non-Hermitian Hamiltonians},}
  {\protect\JournalTitle{Reports on Progress in Physics}} \textbf{70},
  947--1018 (2007).

\bibitem{bender2018pt}
C.~M. Bender, \emph{PT Symmetry: In Quantum and Classical Physics} (World
  Scientific Publishing, 2018).

\bibitem{Mostafazadeh2002c}
A.~Mostafazadeh, \enquote{{Pseudo-Hermiticity versus PT symmetry: The necessary
  condition for the reality of the spectrum of a non-Hermitian Hamiltonian},}
  {\protect\JournalTitle{Journal of Mathematical Physics}} \textbf{43},
  205--214 (2002).

\bibitem{Mostafazadeh2002d}
A.~Mostafazadeh, \enquote{{Pseudo-Hermiticity versus PT-symmetry. II: A
  complete characterization of non-Hermitian Hamiltonians with a real
  spectrum},} {\protect\JournalTitle{Journal of Mathematical Physics}}
  \textbf{43}, 2814--2816 (2002).

\bibitem{Mostafazadeh2002e}
A.~Mostafazadeh, \enquote{{Pseudo-Hermiticity versus PT-symmetry III:
  Equivalence of pseudo-Hermiticity and the presence of antilinear
  symmetries},} {\protect\JournalTitle{Journal of Mathematical Physics}}
  \textbf{43}, 3944--3951 (2002).

\bibitem{Longhi2017a}
S.~Longhi, \enquote{{Parity-time symmetry meets photonics: A new twist in
  non-Hermitian optics},} {\protect\JournalTitle{Epl}} \textbf{120} (2017).

\bibitem{Ruschhaupt2005a}
A.~Ruschhaupt, F.~Delgado, and J.~G. Muga, \enquote{{Physical realization of
  PT-symmetric potential scattering in a planar slab waveguide},}
  {\protect\JournalTitle{Journal of Physics A: Mathematical and General}}
  \textbf{38} (2005).

\bibitem{Ruter2010}
C.~E. R{\"{u}}ter, K.~G. Makris, R.~El-Ganainy, D.~N. Christodoulides,
  M.~Segev, and D.~Kip, \enquote{{Observation of parity-time symmetry in
  optics},} {\protect\JournalTitle{Nature Physics}} \textbf{6}, 192--195
  (2010).

\bibitem{Berry_2008}
M.~V. Berry, \enquote{Optical lattices with {PT} symmetry are not transparent,}
  {\protect\JournalTitle{Journal of Physics A: Mathematical and Theoretical}}
  \textbf{41}, 244007 (2008).

\bibitem{PhysRevLett.106.213901}
Z.~Lin, H.~Ramezani, T.~Eichelkraut, T.~Kottos, H.~Cao, and D.~N.
  Christodoulides, \enquote{Unidirectional invisibility induced by
  $\mathcal{P}\mathcal{T}$-symmetric periodic structures,}
  {\protect\JournalTitle{Phys. Rev. Lett.}} \textbf{106}, 213901 (2011).

\bibitem{nature_12}
A.~Regensburger, C.~Bersch, M.-A. Miri, G.~Onishchukov, D.~N. Christodoulides,
  and U.~Peschel, \enquote{Parity–time synthetic photonic lattices,}
  {\protect\JournalTitle{Nature}} \textbf{488}, 167 (2012).

\bibitem{nature_13}
L.~Feng, Y.-L. Xu, W.~S. Fegadolli, M.-H. Lu, J.~E.~B. Oliveira, V.~R. Almeida,
  Y.-F. Chen, and A.~Scherer, \enquote{Experimental demonstration of a
  unidirectional reflectionless parity-time metamaterial at optical
  frequencies,} {\protect\JournalTitle{Nature Materials}} \textbf{12}, 108--113
  (2013).

\bibitem{Longhi2010}
S.~Longhi, \enquote{{Spectral singularities and Bragg scattering in complex
  crystals},} {\protect\JournalTitle{Physical Review A - Atomic, Molecular, and
  Optical Physics}} \textbf{81}, 1--6 (2010).

\bibitem{PhysRevAaeva}
E.-M. Graefe and H.~F. Jones, \enquote{$\mathcal{PT}$-symmetric sinusoidal
  optical lattices at the symmetry- breaking threshold,}
  {\protect\JournalTitle{Phys. Rev. A}} \textbf{84}, 013818 (2011).

\bibitem{PhysRevA.98.053844}
D.~M. Tsvetkov, V.~A. Bushuev, V.~V. Konotop, and B.~I. Mantsyzov,
  \enquote{Broadband quasi-$\mathcal{PT}$-symmetry sustained by inhomogeneous
  broadening of the spectral line,} {\protect\JournalTitle{Phys. Rev. A}}
  \textbf{98}, 053844 (2018).

\bibitem{PhysRevLett.119.243904}
V.~Achilleos, Y.~Aur\'egan, and V.~Pagneux, \enquote{Scattering by finite
  periodic $\mathcal{P}\mathcal{T}$-symmetric structures,}
  {\protect\JournalTitle{Phys. Rev. Lett.}} \textbf{119}, 243904 (2017).

\bibitem{Shramkova_2016}
O.~V. Shramkova and G.~P. Tsironis, \enquote{Scattering properties {ofPT}-
  symmetric layered periodic structures,} {\protect\JournalTitle{Journal of
  Optics}} \textbf{18}, 105101 (2016).

\bibitem{Wolf2013a}
E.~Wolf, \enquote{{Diffraction of radiation of any state of spatial coherence
  on media with periodic structure},} {\protect\JournalTitle{Optics Letters}}
  \textbf{38}, 4023 (2013).

\bibitem{Gao2018a}
F.~Gao, Y.-M. Liu, X.-D. Tian, C.-L. Cui, and J.-H. Wu, \enquote{{Intrinsic
  link of asymmetric reflection and diffraction in non-Hermitian gratings},}
  {\protect\JournalTitle{Optics Express}} \textbf{26}, 33818 (2018).

\bibitem{Shui2018}
T.~Shui, W.~X. Yang, S.~Liu, L.~Li, and Z.~Zhu, \enquote{{Asymmetric
  diffraction by atomic gratings with optical PT symmetry in the Raman-Nath
  regime},} {\protect\JournalTitle{Physical Review A}} \textbf{97}, 33819
  (2018).

\bibitem{Zhu2016a}
X.~Y. Zhu, Y.~L. Xu, Y.~Zou, X.~C. Sun, C.~He, M.~H. Lu, X.~P. Liu, and Y.~F.
  Chen, \enquote{{Asymmetric diffraction based on a passive parity-time
  grating},} {\protect\JournalTitle{Applied Physics Letters}} \textbf{109}
  (2016).

\bibitem{PhysRevLett.63.2220}
E.~Wolf, \enquote{Correlation-induced doppler-type frequency shifts of spectral
  lines,} {\protect\JournalTitle{Phys. Rev. Lett.}} \textbf{63}, 2220--2223
  (1989).

\bibitem{PhysRevE.52.6833}
M.~Du\ifmmode~\check{s}\else \v{s}\fi{}ek, \enquote{Diffraction of partially
  coherent beams on three-dimensional periodic structures and the angular
  shifts of the diffraction maxima,} {\protect\JournalTitle{Phys. Rev. E}}
  \textbf{52}, 6833--6840 (1995).

\bibitem{wolf2007introduction}
E.~Wolf, \emph{Introduction to the Theory of Coherence and Polarization of
  Light} (Cambridge University Press, 2007).

\bibitem{Brandao2017}
P.~Brand{\~{a}}o and S.~Cavalcanti, \enquote{{Bragg-induced power oscillations
  in PT -symmetric periodic photonic structures},}
  {\protect\JournalTitle{Physical Review A}} \textbf{96} (2017).

\bibitem{berry1998diffraction}
M.~Berry and D.~O'Dell, \enquote{Diffraction by volume gratings with imaginary
  potentials,} {\protect\JournalTitle{Journal of Physics A: Mathematical and
  General}} \textbf{31}, 2093 (1998).

\end{thebibliography}

\bibliographyfullrefs{sample}


\ifthenelse{\equal{\journalref}{aop}}{%
\section*{Author Biographies}
\begingroup
\setlength\intextsep{0pt}
\begin{minipage}[t][6.3cm][t]{1.0\textwidth} 
  \begin{wrapfigure}{L}{0.25\textwidth}
    \includegraphics[width=0.25\textwidth]{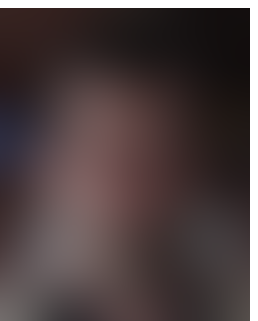}
  \end{wrapfigure}
  \noindent
  {\bfseries John Smith} received his BSc (Mathematics) in 2000 from The University of Maryland. His research interests include lasers and optics.
\end{minipage}
\begin{minipage}{1.0\textwidth}
  \begin{wrapfigure}{L}{0.25\textwidth}
    \includegraphics[width=0.25\textwidth]{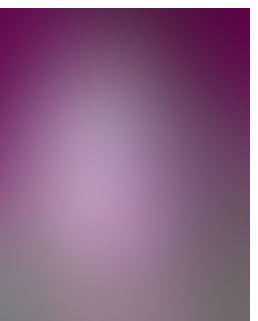}
  \end{wrapfigure}
  \noindent
  {\bfseries Alice Smith} also received her BSc (Mathematics) in 2000 from The University of Maryland. Her research interests also include lasers and optics.
\end{minipage}
\endgroup
}{}

\end{document}